Authors: Vito Veneziano*, Austen W. Rainer** and Sheraz Haider***

*, ** School of Computer Science, University of Hertfordshire
College Lane Campus
AL10 9AB, Hatfield (UK)
*v.veneziano@herts.ac.uk, +44 (0) 1707 284363 (Corresponding author)
**a.w.rainer@herts.ac.uk, +44 (0) 1707 284763

***The IMC Group Ltd
Unit 8-9, Pendle house
SG6 1SP, Letchworth Garden City (UK)
***s.haider@the-imcgroup.com, +44 (0) 1462 688070 ext. 123

Title: When Agile Is Not Good Enough: an initial attempt at understanding how to make the right decision

Abstract: Particularly over the last ten years, Agile has attracted not only the praises of a broad range of enthusiast software developers, but also the criticism of others. Either way, adoption or rejection of Agile seems sometimes to be based more on a questionable understanding than on a critical, well-informed decision making process. In this paper, the dual nature of the above criticism is discussed, and the arguments against Agile have been classified within a critical taxonomy of risk factors. A decisional model and tool based on such taxonomy are consequently proposed for supporting software engineers and other stakeholders in the decision-making about whether or not to use Agile. The tool, which is freely available online, comes with a set of guidelines: its purpose is to facilitate the community of software developers to contribute to further assessing the potential and the criticalities of Agile Methods.

# 1. INTRODUCTION

Software engineering, like any other human activity, is not immune from accepting and adopting approaches and paradigms in a manner that, in other contexts, people wouldn't hesitate to label as "fashionable" and "trendy" those who follow it.

On the same dimension, but with much less levity, adherence to such paradigms can easily become a matter of "orthodoxy", "good practice", "standards".

In the recent past, just to mention one example, object-orientation has established such a culturally dominant paradigm within which to develop software systems, that very few practitioners and academics would feel confident enough in suggesting or pursuing alternative approaches to programming, even when there might be scope for such[1].

Similarly, Agile is currently becoming *the* approach to follow in software engineering.

As the heirs of Toyota Production System (TPS) and lean-manufacturing[2], Agile Methods implement a set of principles and practices, which aim at optimizing efforts and resources whilst attaining fast, effective programming outcomes.

Both the approach and the methods do not come without criticism. Two arguments are usually set out against Agile: the one based on a critical analysis of the several contradictions that are inherent, although often left unacknowledged, in Agile itself, and that aims at depicting it as a myth; and the one grounded on the specific issues that could affect the suitability of adopting Agile for developing given software systems in certain contexts or against certain constraints.

In the next section of this paper, these two kinds of criticism will be considered and discussed in some detail.

In the third section, a number of issues, which are considered to be risk factors, will be grouped in a critical taxonomy. This relates to the second argument against use of Agile.

Next, a hypothetical decisional model based on the critical taxonomy of the identified risk factors is identified. An experimental tool is then derived and defined, aimed at informing practitioners' decisions on whether an agile method is the most suitable for their project. The tool is a questionnaire with weighted answers implementing the decisional model.

In the fifth and final section, after a short description of how such a tool has been used and could lead to the decision not to adopt any Agile method in a specific case-study, some guidelines are provided on how the tool should be used by practitioners for further enhancing its efficacy.

# 2. STATE OF THE ART

## 2.1 Context: Criticism on Agile as a myth

Adoption of Agile has become mainstream in software development (West and Grant, 2010).

This paper proposes that over-reliance on what has become mainstream practice appears to have impaired critical assessment of whether the specific context (a project, a case study, a problem, an organization etc.) would benefit more from a less commonly used approach.

Such an attitude shown by practitioners, whilst –strictly speaking- does not in any way affect the potential of Agile, as the proven effective management approach that it is in a variety of scenarios, still seems to contribute and reinforce what Agile has become overall in terms of common perception: a myth, an ideology, a religion even, whose "evangelists" preach about across many fields and disciplines, well beyond software engineering boundaries.

Myths seem to be ineradicable weeds in the software garden: from what Fred Brooks masterfully described in his "The Mythical Man-Month" (1975) or by his famous paper "No Silver Bullet" (1986), up to the myths of Object-Orientation (James Noble, 2009), which we teach our students in order to "shape the programs they write; myths to shape the way they think about the programs they write".

It should not be ignored that many Agile practitioners have honestly tackled and somehow assessed myths as Agile is concerned: remarkable are the ten myths about Agile as reported from Bill Ives (2008), out of a panel discussion led by Jeff McKenna of Serena with members (and agile practitioners) from Valtech and Serena at the Serena Tag 2008 conference.

---

[1] For a short review of arguments, see Cunningham (2013)

[2] Which led to the so-called "lean software development" approach (Poppendieck and Poppendieck, 2003)

But again, how unsurprising is that evangelists end up supporting their faith?

Also, it is worthy to underline that the term "evangelists" is not used here only to emphasize a critic perspective: many practitioners label themselves this way[3].

Unfortunately, only few have tackled Agile as a myth in itself though, and no-one from the "research" community, whose empirical studies and/or contributions were considered by the well-known systematic review of Tore Dyba and Torgeir Dingsøyr (2008).

For all these reasons, a need to look at other sources in order to support the main goals of this study arose: and for the same reasons another systematic review has not even been attempted here, as in most cases referenced sources are far from being of "high quality research".

Moreover, criticizing Agile as a myth -strictly speaking- does not aim at adding anything new to what is known already about limits and constraints of the Agile approach: besides, even the most enthusiast supporter of Agile could easily admit that Agile is not a panacea.

The expected benefit of exposing a sibylline, cryptic and unctuous nature of the Agile myth is rather to be found in the consequential attempt at alerting stakeholders not to act as acritical supporters nor as prejudicial antagonists: whatever the personal attitude, it should always be possible to decide not to adopt Agile *after* careful consideration of the limits even for the enthusiasts; or, on the contrary, to adopt it if conditions and circumstances would really allow so.

In other words, declaring Agile a myth (to both its supporters and antagonists) helps filling a gap between *what* it is known already and *how* to use such a knowledge in order to make a suitable decision on whether to go Agile or not.

The challenge here is to mitigate (or amplify, when needed) each stakeholder bias or attitude towards Agile (either positive or negative) by means of a strategy that would negotiate the emotional side of that stakeholder against a more neutral evaluative approach. Accordingly to the decisional model we propose in the following sections of the current paper, personal attitudes towards Agile will eventually be translated and implemented as the first weighted question ("Would you describe yourself and your team as a supporter of Agile?"). The subsequent answer to this question, to be given from within a range of decimals between the "Absolutely not" (0) and the "Absolutely yes" (1) values, would somehow play the crucial role of calibrating the overall decisional tool against stakeholders attitude.

Awareness of personal attitudes within a development team seems particularly appropriated when Agile is perceived as a safe option "no matter what": as Steve Yegge (2006) wrote in his well-known "Good Agile, Bad Agile" post in his personal blog, underestimating how deeply the inappropriate adoption of Agile can affect the outcome of a given project is a terrible risk to run, especially when people believe that "Hey, it can't hurt. I'll give it a try".

## 2.2  Context: Criticism on Agile specific issues

In the assumption that practitioners and stakeholders in any given project have agreed that the adoption of an agile method should be the outcome of a lucid decision to be made accordingly to some criteria and not just an axiom to be accepted (or rejected) without controversy, then the next step should be represented by the task to systematically identify and assess which factual issues (which should be considered as *risk factors*) about Agile suitability are to be considered relevant within the economy of that specific project.

Literature provides already a rich collection of issues concerning the suitability of adopting agile processes in software engineering contexts: they come under the form of limitations, criticalities, inconsistencies, etc.

It must be underlined that each one of these issues has been the object of a perennial litigation between both the parties (Agile supporters VS antagonists). In one post of his blog, Michael Dubakov (2007) highlighted how "erroneous and dangerous" can be a certain type of criticism on Agile, mostly based on a superficial assessment and unsuitable comparison of possibly irrelevant cases or applied to circumstances where criticalities were not inherent the adoption of an Agile methodology. This means that such a list of issues should always be considered under a dialectical perspective:

- on one side, for mitigating any excess of criticism, as the issues tackle generic scenarios which could occur in or refer to highly different contexts, it is important to ensure that these differences are duly taken into account for each specific case;

---

[3] On Google (UK Edition), about 76,400 results were obtained by searching for "agile + evangelist", on the 12 July 2012, and as expected the vast majority were linking to infomercials from Agile consultants and practitioners websites, blogs, etc.

- on the other side, in order to avoid the temptation of neglecting any criticism on the basis of their supposedly weakness, it is important to remember that issues often originate from development cases in which, notwithstanding (or maybe just because of) the adoption of Agile methods, problems and/or failures have nevertheless been reported, thus creating an associative record between problems/failures and Agile Methods. In the most benevolent interpretation, it is safe to claim that Agile didn't help preventing a failure even if this was caused or determined by some other circumstances.

For the purposes of this study, most of mentioned issues have been gathered from a paper that Turk et al. (2002) presented at XP2002. In that paper, the authors argued that eleven underlying assumptions had to be considered valid (i.e., granted by the project circumstances) if any benefit from the adoption of an agile process was to be expected.

Shortly, these assumptions were:

1. Co-location, or at least full and prompt availability, of all development stakeholders, including customer representatives, is a must
2. Documentation and modeling are expendable outcomes of the development process
3. Volatility of requirements within an evolving environment are both inescapable facts
4. Adopting development processes that better adapt to changing projects are more likely to outcome high-quality products
5. Within an Agile organization, developers personal competences and skills are sufficient for supporting the adapting of the development process to the evolving project
6. An incremental approach to development ("and a few metrics") is the main strategy to follow for developing software
7. Evaluation of software is mostly performed by informal reviews and code testing
8. Reusability is out of scope when developing application-specific software
9. Development costs do not significantly increase "just because" of any occurring change
10. Given certain circumstances, any software can be developed in increments
11. There is no need to design for change because code-refactoring is all it is needed for handling any change

From the above assumptions, Turk et al. derived and discussed a list of limitations which have been fully considered and included in the following critical taxonomy (specifically, issues no. 04, 05, 06, 07, 08, 09, 10, 11 and 12), together with other issues that other sources have highlighted or that the Authors of the current paper have suggested themselves on the grounding of their personal experience.

## 3. A CRITICAL TAXONOMY OF ISSUES (OR RISK FACTORS) ON AGILE

A critical taxonomy of 19 issues has been developed. Each issue (or Risk Factor, R$n$) should be read as coming under the form of a question beginning with "How important/impacting is that…". For example, issue no. 2 ("Documentation cannot be considered and thoroughly prepared as a critical asset") should be read as "How important is that documentation cannot be considered and thoroughly prepared as a critical asset?"

A threefold categorization of issues or risk factors has been assumed: the first and most populated category includes issues that are inherent to some intrinsic features of Agile (issues 1, 2, 3, 4, 5, 6, 7, 8, 9, 10, 11, 13, 14, 16, 18). Other risk factors (issues no. 12 and 15) directly emanate from the view that Agile can be seen as a "money-making exercise for a group of consultants" (Halliwell, 2008; Yegge, 2006), which –when any conflict of interested is proven true- could obviously come with some nasty side-effects. A third category of issues (issues no. 17 and 19) would relate to processes other than development, and could challenge stakeholders to ensure that those processes are under control and performed honestly by the team.

Table 1. A critical taxonomy of potential issues (or risk factors) on adopting Agile Methods

| R$n$ | Description of the Issue (or Risk Factor) |
|---|---|
| 01 | The customer representative cannot be always available and present alongside the development process |
| 02 | A final "weaker" (e.g., less complete than expected) User Acceptance Test is likely |

| 03 | A less reliable initial prediction on time and budget is to be expected |
|----|---|
| 04 | Documentation cannot to be considered and thoroughly prepared as a critical asset (Turk et al., 2002) |
| 05 | An agile methodology, with its set of practices, as such, can be less flexible than expected by an agile approach overall (Turk et al., 2002; Yegge, 2006; Halliwell, 2008) |
| 06 | Little experience and somehow more relaxed discipline in Agile could reflect negatively on project management (Turk et al., 2002) |
| 07 | Limited or unsustainable support for distributed development environments is likely to occur (Turk et al., 2002) |
| 08 | Outsourcing and/or subcontracting is likely to be more difficult to manage (Turk et al., 2002) |
| 09 | Limited support and/or opportunities for building reusable artifacts is likely to occur (Turk et al., 2002) |
| 10 | Limited or unsustainable support for development involving large teams is likely to occur (Turk et al., 2002) |
| 11 | Limited supported for developing safety-critical software is likely to occur (Turk et al., 2002) |
| 12 | Limited support for developing large, complex software is likely to occur (Turk et al., 2002; Dybå and Dingsøyr, 2008) |
| 13 | The lack of focus on architecture is bound to engender suboptimal design-decisions (Dybå and Dingsøyr, 2008) |
| 14 | Because of social values embraced by Agile, decision making can be less effective than expected (Dybå and Dingsøyr, 2008) |
| 15 | Agile developers/consultants might not be around for long (Halliwell, 2008) |
| 16 | Agile methods could tackle only the "trivial" part of a project and leave behind the tricky, hard ones (Halliwell, 2008) |
| 17 | Development process and outcomes could depend on the quality of the hiring process (people are more important than processes and tools) (Halliwell, 2008) |
| 18 | Adaptive planning could be "practically" (for instance, when responding to a change) translated into no long-term planning (Halliwell, 2008) |
| 19 | Planning poker, as a variation of Wideband Delphi (Boehm, 1981), and other distributed decision-making strategies could be affected (both unwillingly or otherwise) by individuals who are not focused on the actual development[4]. |

Whilst the content of Table 1 is arguably far from being exhaustive, it should include a list of issues that most practitioners might find familiar to their own experience.

In particular, issue no. 5 seems significantly relevant to those development teams with skilled and well-matched individuals, whose practice is already oiled and could barely benefit from some extra codified set of practices (Halliwell, 2008: see Ryan Cooper in the comments).

## 4. THE DECISIONAL MODEL

The basic rationale behind the possibility to justify the development of a decision-making model is that each issue or risk factor $r$, weighted against a suggested value $w$ (which the team should agree on, in a range of decimals between 0 and 1), would contribute towards the assessment of the overall risk for adopting an agile method. The Overall Specific Risk (OSR), comprised between 0 and 1, could be defined asthe sum of the $n$ risk factors associated to each issue from the critical taxonomy, and should return the main –but yet unmitigated- decisional value on the overall risk of adopting Agile in a given project.

1. $\text{OSR} = \frac{\sum_{k=0}^{n} w_k r_k}{n}$

The Mitigation-Amplification Factor (MAF) is the value associated to the team attitude to Agile (as a myth) and used to calibrate the decisional value derived by the OSR.

---

[4] See, for example, the so-called anchoring effect (Tørresdal, 2007)

Given an attitude value towards Agile (hereby referred as AVA), measured against the whole development team and comprised between 0 and 1 (where 0 would mean an extreme critic of Agile and 1 an extreme supporter), MAF is calculated as follow:

2. $\text{MAF} = \log\left(\frac{0.5 + \text{AVA}}{1.5 - \text{AVA}}\right) * \min(|\text{OSR}|, |1 - \text{OSR}|)$

The MAF component comes with the typical inverse hyperbolic tangent curve, which becomes steeper towards the extremes of the (-1,1) interval, which would respectively represent the most extreme attitudes: in our model, the curve has been slightly translated (1±0.5) to better accommodate the values of MAF within the expected range to be used for the definition of the final decisional value (DEC) to be comprised between 0 and 1 (and easily converted into a percentage format).

The final decisional value is defined by the algebraic sum of the overall specific risk plus the mitigation factor (which could be a negative or positive value, depending on whether the team attitude is more or less Agile-oriented, or biased, than an ideal neutral view, represented by an AVA value of 0.5), and would be defined as follows:

3. $\text{DEC} = \text{OSR} + \text{MAF}$

The decisional model could even allow each risk factor to be considered as the holder of an intrinsic weight (the actual risk value) different from 1 (but always $0 \leq r \leq 1$), which impact can be further diminished by its weighting. But, for usability purposes, during our research it has always been assumed that $r = 1$.

We have assumed that a value of DEC > .5 (i.e., greater than 50%) is the threshold over which the adoption of any agile method could be overly risky. This assumption needs obviously to be further validated by empirical evidence: hence, the invitation moved to practitioners to test the model (by means of the tool as described in the following section) and to report on any findings to the corresponding author of this paper.

## 5. THE TOOL, THE APPLICATION CASE STUDY AND USER GUIDELINES

The tool, in its first release, has been called WAINGE, from the initials of the title of this paper, which in turn had been the starting question of the current study. WAINGE has been developed in MS Excel, and comes with a GUI as represented in Figure 1. WAINGE is downloadable from the url provided by Veneziano (2012).

Usability of WAINGE is quite straightforward: users who might like to test it in order to support their decision-making should input the weights for each risk factor as perceived from within their project (in the "Weights" column), and give a value to their team attitude towards Agile in the AVA field.

The DEC output is represented on the bottom right of the spreadsheet.

Figure 1. WAINGE tool v.1.8 GUI

WAINGE has been adopted and tuned for the first time within the context of a Knowledge Transfer Partnership (or KTP) funded project between the University of Hertfordshire and the IMC Group.

The main goal of the project was the development of a web enabled data capture & reporting software application for managing industrial field-based sensors.

At least three problems originating from and impacting many aspects of the IMC business operations and efficacy were tackled by the development of such a software system:

1. IMC customers were relying on traditional standalone software applications, installed by their own premises, for recording and monitoring data (saved and managed by flat-files) such as temperature, humidity etc. from the in-field-mounted instruments they've bought from IMC. This gave rise to multiple sites using independent, often over-customised systems in isolation with no overall visibility of performance or trends.

Solution: The new software system was expected to successfully solve this problem by saving and managing the sensors data on a central database system and by coming with a web based front-end interface, accessible from anywhere on the Internet, subject to several degrees of authorisation from both IMC and/or the licensee customer.

2. IT departments of large companies often change systems and configurations without regard to their supplier interfaces. This caused IMC Group many and expensive management and maintenance issues, as three support engineers were spending about 80% of their time by just addressing IT problems at local sites.

Solution: The new software system was expected to reduce the amount of support on IT issues down from multiple sites (up to hundreds) to virtually one. Better resource utilisation of support engineers would create efficiencies in labour and travel costs. The opportunity of providing access to needed data by means of a web-enabled system at one site would also open up new markets for remote monitoring such as energy consumption. The assumption here was that, ultimately, customers are interested in their data, not the instruments.

3. IMC Group quoted at the start of this KTP project "The competition is moving in the direction of web enabled access, so it is imperative we do not get left behind". To achieve this expertise in developing highly usable and easily accessible graphical representations of real-time data for customer access was needed.

Solution: The new software system was expected to solve this problem as well, by taking full advantage of the web technology: by hosting servers which are secure, reliable, backed up and becoming more acceptable to industry, by developing server-based applications which are easier to maintain and upgrade and by adopting standards for supporting the deployment of complex and yet usable web-based GUIs.

Suitably addressing the above problems was critical to the success of the project, so it was decided by the KTP team to systematically refer to these problems in order to eventually provide the matrix of weights to be used in the WAINGE tool.

The KTP team consisted of a number of stakeholders:

1. The academic coordinator (a lecturer and a consultant in software engineering)
2. The academic supervisor (a lecturer and a consultant in software engineering)
3. The academic associate (a recent MSc graduate in software engineering)
4. The company technical director (a physicist and a software designer, who also developed the existing standalone software systems)
5. The company supervisor (an electrical and signal engineer)

Each stakeholder had a different grip on and exposure to software development and agile methods, with at least one team member recording a professional experience in an Agile method development (namely, FDD): thus, in principle, nothing would have prevented the KTP team as a whole to adopt an agile method, if that was the case.

However, the team agreed to make a final decision by using the WAINGE tool, and some simple guidelines were followed.

The first and more important guideline was to match the above mentioned three problems with/against each risk factor of the critical taxonomy.

For example, risk factor no. 1 ($R_{01}$: "The customer representative cannot be always available and present alongside the development process") was considered against the three problems, which in turn were summarized by three shorter statements, as listed below:

$P_1$: Moving the software out of customer premises and reducing the amount of expected customization by the supplier (IMC)

$P_2$: Improving the maintenance processes and making the software less platform-sensitive

$P_3$: Supporting customers in managing and accessing from everywhere their front-end GUI without affecting IMC overall control on server-side main functionalities and authorization processes.

In order to use the risk factor against the three statements, three questions were formulated as conditional statements, as follows:

$Q1_{r1}$: if $R_{01}$, then how this could affect $P_1$

$Q2_{r1}$: if $R_{01}$, then how this could affect $P_2$

$Q3_{r1}$: if $R_{01}$, then how this could affect $P_3$

In our example, $Q1_{r1}$ ("If the customer representative cannot be always available and present alongside the development process, then how this could affect our moving the software out of customer premises and reducing the amount of expected customization by the supplier?") attracted a weight value of 1 (very high impact), Q2 attracted a weight value of 0.5, Q3 attracted a weight value of 0.9: overall, the outcome weighting for $R_{01}$ was the resulting average 0.8 (i.e., $w_{01} = 0.8$).

The full matrix of weights for the 19 risk factors considered has been reported as an annex.

The second guideline concerned the allocation of the AVA (attitude towards Agile) value: after a short introduction to the Agile manifesto and a zoom on two methods (SCRUM and FDD), each team member was asked to reflect and self-assess their own attitude, on the groundings of their own experience and on the discussion that followed the introduction to the Agile manifesto and methods. The final value was the simple average calculated between the whole KTP team and resulted into a cumulative slightly biased attitude against Agile (AVA = 0.4). It has to be underlined that, accordingly to our model, such a value would somehow "push" the final Decision value towards Agile.

Before the actual calculus, as a third and final guideline it had been agreed to identify which candidate methods the team would have chosen to adopt. The Agile method was FDD; the non-Agile was the Spiral Model, which was still an iterative approach. An iterative approach was chosen over phased development because the development team could anticipate that many new requirements would emerge alongside the development process, and they were easier to accommodate than from within a more traditional development process model such as the Waterfall process model. Also, the nature of this project was such that its scope would change (and possibly become broader and broader) as new market conditions and expectation arise.

Given the final DEC value (0.576, or 57.6%), the Spiral Model was adopted.

The KTP project lasted 18 months and ended successfully in 2011. Not only the software was eventually developed up to expected standard and requirements, but also, at the end of the KTP, the academic associate has been employed by the Company.

This is far from claiming that the tool was accurate in predicting that Agile should have not been adopted, and as stated before more investigation and empirical evidence is needed. But, independently from its mathematical decisional power, it is safe to assume that the tool (or the systematic approach behind it in considering all issues associated with Agile) helped the development team to focus on what was important and how each factor could have affected the overall outcome.

## 6. CONCLUSIONS

A critical taxonomy of issues, which are known to be associated to the adoption of agile methods, has been identified and developed in order to help deciding whether to go or not to go Agile, given some specific circumstances (contexts, problems, goals, etc.) inherent a given software development project.

The taxonomy, used in conjunction with a decisional tool called WAINGE, provided the authors of this paper with the opportunity to evaluate in detail and systematically whether Agile was a suitable option and, without excluding some luck, the made decision was supported by a successful outcome of the project.

In this paper, some detailed information about the decisional model together with some simple guidelines used on how to systematically address relevant issues in order to weight each risk factor, have been provided.

Whilst more empirical evidence is needed (hence our encouragement to practitioners and researchers to adopt the tool and send their findings to the corresponding author), it is a reasonable expectation that the overall community of practitioners would benefit from a more lucid and critical approach to Agile (and to its criticism as well): and we are confident that our decisional model (or at least the systematic evaluative approach behind it) would provide some promising hints in this direction.

## 7. ACKNOWLEDGMENTS

This paper and the research behind it has been partially funded by the KTP Programme No. 7593, granted to the University of Hertfordshire, Hatfield, UK, and to the IMC Group, Letchworth Garden City, UK

## 8. ANNEX: THE MATRIX OF WEIGHTS AND THE *AVA* ADOPTED IN THE KTP

Matrix of risk values and context-sensitive weights specifically considered for the projectas perceived by the development team for each risk: when no agreement was achieved by the team, the weights were given a neutral value of 0.5.

All the risks carry a default risk value of 1, even when they are not explicitly grounded in literature: it is left up to the teamto find an agreement on how to weight them.

At the time of the decision, the case-study team was not significantly biased towards or against Agile: however it was decided to allocate the ATA parameter a value of 0.4 (slightly unbalanced against Agile).

Table 2: Matrix of weights per risk factor

| *Rn* | Description | *w* |
| --- | --- | --- |
| 01 | The customer representative cannot be always... | 0.8 |
| 02 | A final "weaker" (e.g., less complete than... | 0.8 |
| 03 | A less reliable initial prediction on time and... | 0.7 |
| 04 | Documentation cannot to be considered and... | 0.8 |
| 05 | An agile methodology, with its set of practices... | 0.6 |
| 06 | Little experience and somehow more relaxed... | 0.5 |
| 07 | Limited or unsustainable support for distributed... | 0 |
| 08 | Outsourcing and/or subcontracting is likely to... | 0 |
| 09 | Limited support and/or opportunities for... | 1 |
| 10 | Limited or unsustainable support for... | 0 |
| 11 | Limited supported for developing safety-critical... | 0.5 |
| 12 | Limited support for developing large, complex... | 0.8 |
| 13 | The lack of focus on architecture is bound to... | 1 |
| 14 | Because of social values embraced by Agile... | 0.6 |
| 15 | Agile developers/consultants might not be... | 0.9 |
| 16 | Agile methods could tackle only the "trivial"... | 0.8 |
| 17 | Development process and outcomes coud... | 0.7 |
| 18 | Adaptive planning could be "practically" (for... | 0.3 |
| 19 | Planning poker (as a variation of Wideband... | 0.8 |